\documentclass[aps,prl,twocolumn,showpacs,preprintnumbers,amsmath,amssymb]{revtex4}

\usepackage{graphicx}
\usepackage{dcolumn}
\usepackage{amsmath,mathrsfs,amssymb,amsthm}
\usepackage{hhline}
\usepackage{wasysym,enumerate,color}
\usepackage{epstopdf}
\usepackage{verbatim}
\usepackage{hyperref}
\usepackage{color}
\usepackage{ulem} 
\renewcommand{\emph}[1]{\textit{#1}} 

\definecolor{darkgreen}{rgb}{0,0.5,0}
\definecolor{purple}{rgb}{0.35,0,0.35}
\definecolor{orange}{rgb}{1,0.5,0}
\definecolor{darkred}{rgb}{.7,0,0}
\definecolor{darkblue}{rgb}{0,0,.3}
\definecolor{grey}{rgb}{.6,.6,.6}
\definecolor{dimgreen}{rgb}{0.2,0.6,0.1}
\hypersetup{colorlinks,linkcolor=blue,urlcolor=blue,citecolor=blue}

\newcommand{\be}{\begin{equation}}
\newcommand{\ee}{\end{equation}}
\newcommand{\bea}{\begin{eqnarray}}
\newcommand{\eea}{\end{eqnarray}}

\newcommand{\calS}{{\cal S}}

\newcommand{\e}{\varepsilon}
\newcommand{\sgn}{{\rm sgn}}

\newcommand{\s}{\sigma}

\newcommand{\mycomma}{, }

\begin{document}

\title{Fermi liquid theory of resonant spin pumping}
\author{  C. P. Moca,$^{1,2}$     A. Alex,$^{3}$ 
A. Shnirman$^{4}$, and G. Zarand,$^{1}$}
\affiliation{
$^1$BME-MTA Exotic Quantum Phases "Lend\"ulet" Group, Institute of Physics,
Budapest University of Technology and Economics, H-1521 Budapest, Hungary
\\
$^2$ Department of Physics, University of Oradea, 410087, Oradea, Romania
\\
$^3$  Physics Department, Arnold Sommerfeld Center for Theoretical Physics and Center for NanoScience,
Ludwig-Maximilians-Universit\"at M\"unchen, D-80333 M\"unchen, Germany
$^4$  Institut f\"ur Theorie der Kondensierten Materie\mycomma Karlsruhe Institute of Technology, 76128 Karlsruhe, Germany 
}
\date{\today}

\begin{abstract}
We study resonant all-electric adiabatic spin pumping through a quantum dot with two nearby levels by using a Fermi liquid approach 
in the strongly interacting  regime, combined with a projective numerical renormalization group (NRG) theory. 
Due to  spin-orbit coupling, a strong spin pumping resonance emerges at every  charging transition, which allows 
for the transfer of  a spin  $\sim \hbar/2$ through the device in a single pumping cycle. 
Depending on the precise geometry of the device, controlled pure spin pumping is also possible. 
\end{abstract}


\maketitle

\emph{Introduction:} 
Spin-orbit (SO) coupling plays a prominent  role in many different fields of physics: it  is not only 
responsible for magnetic anisotropy 
and thus determines the orientation  and low energy excitation spectra of magnets 
and magnetic molecules~\cite{ReviewOnMagnetism}, but its presence also  changes the universality 
class of the localization transition~\cite{LocalizationReview}, and it is also a crucial component for realizing topological 
insulators~\cite{Molenkamp,BrouwerMajorana,OppenMajorana}. The SO coupling  plays also a determining role in mesoscopic physics, 
in spintronics, and, most importantly, in spin-based  quantum computation.  In the latter context, in particular,  it produces spin 
relaxation in spin quantum bits~\cite{Khaetskii,Loss} and leads to geometrical spin relaxation even in the absence of 
external magnetic fields \cite{SanJose}, however,  it  can also be used to 
generate  effective magnetic fields and achieve electrical spin control ~\cite{Nazarov}. 

It has been first  observed in Ref.~\cite{BrouwerPumping} that, in the presence of SO interaction, 
  one can produce a  spin current  by simply cycling adiabatically the parameters of 
a chaotic cavity  (pumping) without breaking the instantaneous time reversal  symmetry, i.e., without applying an external 
magnetic field. Obviously, realizing such spin pumps would enable one to reach an important goal of spintronics, and 
build all electric spin sources. Indeed, guided by this observation, more controlled setups have been proposed to pump spin currents 
through quantum wires~\cite{Fazio} and quantum dots~\cite{Brosco}, however, the effects of interactions 
were ignored in all these studies. While this is justified to a certain extent for the case of a 
quantum wire~\cite{Fazio}, it is certainly unjustified for a quantum dot~\cite{Brosco}, where --
precisely in the regime of interest -- interactions are necessarily strong~\cite{Brosco}.
Studying pumping through  strongly correlated systems is a notoriously hard problem~\cite{Citro}. For charge pumping through 
quantum dots,  several expressions have been derived based upon an adiabatic expansion of the 
Keldysh Green's functions~\cite{Sela,Fazio}. The expressions obtained, however, contain terms, which correspond to local 
charge oscillations, not related to true pumping. An alternative, perturbative approach of pumping 
has been developed in Ref.~\cite{Konig}, but this method is restricted to the regime of weak tunneling and high 
temperatures, and cannot be used to reach the most exciting low temperature regime. 

Here we revisit the problem studied in Ref.~\cite{Brosco} and investigate how the interplay of  
SO coupling and \emph{ strong electronic interactions}  influences spin pumping through a quantum dot at 
very low temperatures,  deep in the strongly correlated regime. 
Our method is very different in spirit from those of  Refs.~\cite{Fazio,Sela,Konig},  and rather, it follows lines 
similar to Ref.~\onlinecite{Aono}: we start out from the 
observation that at $T=0$ temperature our quantum dot (similar to many interacting systems of interest)  realizes a 
local Fermi liquid state. In this state, quasiparticle scattering at the Fermi energy is elastic, and 
can be characterized by a single particle on shell S-matrix. For very small pumping frequencies
 and  small temperatures, $\omega,T\to 0$, 
the current through the device is carried by quasiparticles at or very close to the Fermi surface, where -- to leading order --
multiparticle  scattering processes can be neglected by simple  Fermi liquid phase space arguments. 
Then for the dominant elastic processes, Brouwer's pumping formula can be applied, and the leading contribution 
to the pumped current  can be expressed just in terms of  the single particle 
S-matrix, evaluated at the Fermi energy.   This adiabatic Fermi liquid approach is justified as long as 
$\omega$ and $T$ are less than the Fermi liquid scale (i.e.  the level width  $\Gamma$
in the mixed valence region considered here).

\begin{figure}[t]
\includegraphics[width=0.85\columnwidth,clip]{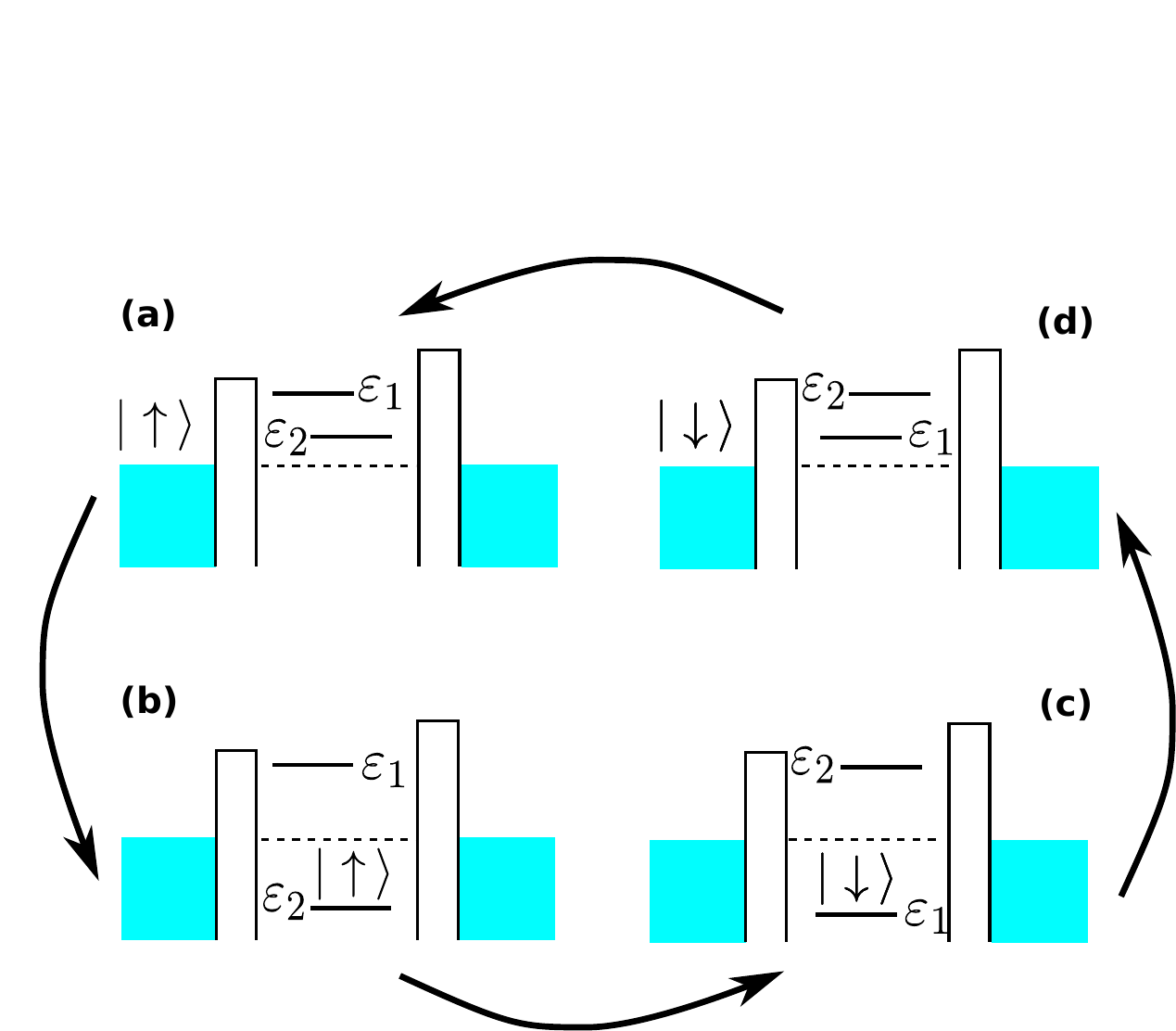}
\caption{\label{fig:pumping_cycle} (Color online)
Sketch of the spin pumping cycle. 
In the initial configuration 
one electron is injected into the dot. Due to the spin orbit
interaction the spin up and spin down parts of the wave function 
are rotated differently (for clarity, only the spin up component is shown), and the spin up and spin down parts 
move into differnt electrodes, thereby resulting in spin pumping.}
\label{fig:sketch}
\end{figure}

Computing the latter 
is still an extremely demanding task: we do that here 
in the most interesting 
narrow level limit by using a projective approach, whereby we first project the Hamiltonian to the 
subspace of Kramers degenerate levels participating in the pumping cycle, and then perform 
numerical renormalization group (NRG) calculations for this projected Hamiltonian and 
reconstruct the  S-matrix.  The strong Coulomb repulsion has a dramatic effect: in the vicinity of 
every charging transition, a spin pumping resonance (or antiresonance) emerges. As a consequence of large  
 interaction,  these resonances are well separated in parameter space, and the total spin pumped through 
them can reach values of $\sim \hbar/2$ in pumping cycles sketched in Fig.~\ref{fig:sketch}. 
 These findings must be contrasted with the hight temperature results of \cite{Konig} and also 
our non-interacting results, Ref.~\cite{Brosco}, where positive and negative pumping regions 
were found to appear close together in  parameter space.

\emph{Model.}
We consider an interacting system  with two almost degenerate 
levels, $\varepsilon_1$ and $\varepsilon_2$, close to the Fermi energy, and weakly coupled to external electrodes.  
The average energy $\bar \varepsilon = (\varepsilon_1+\varepsilon_2)/2$
as well as the energy difference $\Delta\varepsilon = \varepsilon_1-\varepsilon_2$ of these levels can be tuned by  applying several gate voltages to the same quantum dot (or chaotic cavity), as in the experiments of Ref.~\onlinecite{GoldhaberST}. We  shall 
thus consider these as pumping variables throughout this  paper
[$\bar \varepsilon =  \bar \varepsilon (t)$ and  $\Delta\varepsilon = \Delta\varepsilon (t) $].
By disregarding the other occupied or empty levels, we  describe our system 
by the following Hamiltonian
\begin{eqnarray}
H &= &  \sum_{\s,j=\{1,2\}} \e_j(t)\, d^{\dagger}_{j\s} d_{j\s}
+\sum_{\s}\bigl(  t_\s d^{\dagger}_{1\s}\,  d_{2\s}+h.c. \bigr)
\nonumber
\\
&+&  \frac U2 n(n-1) + \sum_{ \s, j, r} v_{j}^{r}\left( d^{\dagger}_{j\s}\psi_{r\s}+h.c.\right )  
\,,
\label{H_dot}
\end{eqnarray}
with $d^{\dagger}_{j\s}$  the creation operator of a spin $\s$  electron at level $j=1,2$, and 
$n$ the total number of electrons on the dot. For simplicity, 
we have chosen the spin quantization axis to coincide with the one dictated by the SO coupling, but 
otherwise  assumed the most general single particle Hamiltonian allowed by time reversal 
symmetry. The parameters  $t_\s =  t+i \alpha \,\s$ 
describe spin dependent hybridization between the two levels  with $\alpha$ the effective strength 
of the SO interaction. The term $\sim Un^2$ accounts for electron-electron interaction, 
 while the last term of Eq.~\eqref{H_dot}  describes the hybridization between the dot levels 
and the leads. The field $\psi^\dagger_{r\s}=\sum_{\mathbf k} c_{{\mathbf k},r\s}^\dagger/(\varrho_r)^{1/2}$ 
creates a conduction electron  in lead $r=L/R$~\footnote{In the regime studied here Hund's rule coupling plays no essential role.},
 and has been normalized by the density of states of the corresponding electrode, $\varrho_r$ so that the hopping amplitudes
 $v_{j}^{r}$ are dimensionless.
 We shall assume that the leads behave as regular Fermi liquids, and thus the 
dynamics of  the creation operators $c_{{\mathbf k},r\s}^\dagger$ (and those of $\psi^\dagger_{r\s}$) are governed by free electron Hamiltonians.


The last term of Eq.~\eqref{H_dot} induces quantum fluctuations and a finite but asymmetrical  broadening of the 
two levels. In the mixed valence regime discussed here, all energy scales must be compared to the strength of these quantum fluctuations, $\Gamma \equiv \sum_{i}\Gamma_{ii} \equiv  2\pi\sum_{r=L/R}\, {v^{r}_{i}}^\ast\,v^{r}_{i}$, 
which  shall be used in what follows  as an energy unit. The hybridization $v^r_i$ induces spin pumping as long as 
the two levels do not couple to the same linear combination of the leads, $\mathrm{det}(v^{r}_{i})\ne 0$ \cite{Brosco}.

\emph{Formalism:} 
As shown by Brouwer~\cite{BrouwerPumping}, for a noninteracting mesoscopic system, for adiabatical parameter changes, the accumulated charge and spin depend only on the path followed in parameter space, and  can both be expressed in terms of the scattering matrix $\calS_{rr}^{\sigma\sigma'}$. Performing  a cycle 
of area $A$ in the parameter space spanned by  $\epsilon_1$ and $\epsilon_2$, e.g.,  one accumulates a spin 
\begin{equation}
\Delta {\mathbf S}_r  =  \frac{\hbar}{2\pi} \int_{A} 
{\boldsymbol \Pi}^{(S)}_{r} (\epsilon_1, \epsilon_2) \,{\rm d}\epsilon_1   {\rm d}\epsilon_2
\label{eq:brower}
\end{equation}
in electrode $r$, where the spin pumping field is defined as 
\begin{equation}
{\boldsymbol \Pi}^{(S)}_{r} (\epsilon_1, \epsilon_2)  =  
{\rm Im\,  Tr} \left \{
\left (
\Lambda_{r}\otimes {\boldsymbol \s}
\right)
\frac{\partial \calS}{\partial \epsilon_2} \frac{\partial \calS^{\dagger}}{\partial \epsilon_1}\right \}\;,
\label{eq:field}
\end{equation}
with $\Lambda_r$ a projector selecting scattering channels in electrode $r$.  As explained in the introduction, 
here we shall  exploit the fact  that the ground state of  Eq.~\eqref{H_dot} is a Fermi liquid~\cite{Nozieres}. Therefore 
quasiparticles scatter elastically at $T\approx 0$, and their scattering  can be described in terms of the 
single particle (on shell) S-matrix evaluated at the Fermi energy, $ S_{rr}^{\sigma\sigma'}(\omega=0)$.
Since precisely these quasiparticles are responsible for adiabatic pumping,  we can continue using   \eqref{eq:brower} at 
very low temperatures, while replacing the noninteracting S-matrix in Eq.~\eqref{eq:field} by its many-body counterpart, $ {\calS}\to S(\omega=0)$. For our Hamiltonian, 
the latter  can be simply related to the Fourier transform of the local  Greens's functions~\cite{Borda}, $G_{j\s,j'\s'} (t)
\equiv -i \langle [d_{j\s}(t) , d^\dagger_{j\s'}(0)] \rangle \theta(t)$,
\begin{equation}
S_{rr'}^{\s\s'}(\omega) = \delta_{rr'} \delta_{\s\s'}
- 2\pi i \sum_{j,j'}\, v^{r}_{j}\,{v^{r'}_{j'}}^\ast\,  G_{j\s,j'\s'} (\omega).	 
\label{eq:FisherLee}
\end{equation}

Our task is thus reduced to compute $G_{j\s,j'\s'} (\omega)$ very precisely as a function of external parameters, and 
then compute the pumped spin. This, however, turns out to be a very challenging task since we need to determine with high precision 
both the imaginary and the real parts of $G_{j\s,j'\s'} (\omega)$ at the Fermi energy. Unfortunately, as of to date, none of the available methods 
can do that reliably.  

Restricting ourself to the most interesting regime of a narrow resonance, $\sqrt{t^2 + \alpha^2} \gg \Gamma$, 
however, we can considerably simplify the problem. For $U=0$ the isolated dot has two Kramer's doublets at energies $E_\pm=\bar \varepsilon \pm \sqrt{t^2 + \alpha^2 + \Delta\varepsilon^2/4}$. Since  $E_+ -E_-\ge 2 \sqrt{t^2 + \alpha^2 }
\gg \Gamma$, for  occupations, $\langle n\rangle \le 2$ we can neglect the higher Kramers doublet, and project to the lower  level, 
$E_-$. 
We thus introduce the operators
\begin{equation}
D^\dagger_\sigma \equiv \sum_j \Phi_{j,\sigma} d^\dagger_{j,\sigma}\,,
\end{equation}
with the spinors $\Phi_\sigma$ parametrized most conveniently in terms of the angles $\varphi\equiv -\cot^{-1}(t/\alpha)$
 and $\vartheta\equiv -\cot^{-1}(\Delta\epsilon/2\sqrt{t^2+\alpha^2})$ and expressed as 
$\Phi_{\uparrow} = \Phi_{\downarrow}^\ast = (\cos(\vartheta/ 2),e^{-i \varphi} \sin(\vartheta / 2) ) $.
The projected Hamiltonian is then just an ordinary Anderson Hamiltonian 
\begin{eqnarray}
H_{\rm proj} &= &  \sum_{\s} E_-(\vartheta,\varphi)  D^{\dagger}_{\s} D_{\s}+  
\frac U2 n(n-1) 
\nonumber 
\\
&+& \tilde v(\vartheta,\varphi) \sum_{ \s} 
\left( D^{\dagger}_{\s}\tilde \psi_{\s}+h.c.\right )  
\,,
\nonumber
\end{eqnarray}
with the hybridization defined as $\tilde v^2 = \sum_r \bigl| {\tilde v}^r_\uparrow \bigr|^2 $, 
with $\tilde v^r_\uparrow \equiv \sum_j \Phi_{j,\uparrow}^\ast v_{j,\uparrow}^r$. Within this approximation, the 
S-matrix of the original fields $\psi_{r,\sigma}$ can then be expressed as 
\begin{equation}
S_{rr'}^{\s\s'}(\omega) = \delta_{\s\s'} \, \bigl\{ \delta_{r,r'}
- 2\pi i \, \tilde v_\s^{r}\,{{\tilde v}_\s^{r'\ast }} \,  G_{D} (\omega) \bigr\}, 	 
\label{eq:Ssimple}
\end{equation}
with $G_{D} (\omega)$ the effective Anderson model's local retarded propagator. At $T,\omega \to 0$  this S-matrix has two eigenvalues for both spin directions: a trivial eigenvalue, $s=1$, and 
an eigenvalue $s=e^{2i\delta}$, with the phase shift $\delta$ related to the occupation of the level $E_-$ by the Friedel sum rule, 
$\langle D_\s^\dagger D_\s\rangle = \delta/\pi$. The   occupation $\langle D_\s^\dagger D_\s\rangle$ is a universal function of the ratios 
$\tilde \Gamma /U$ and $E_- /U$, with $\tilde \Gamma = 2\pi \tilde v^2$ denoting the width of the level $E_-$,
and can be determined reliably by  functional or numerical renormalization group methods as well as by Bethe Ansatz. 
Together with Eq.~\eqref{eq:field}, Eq.~\eqref{eq:Ssimple} thus provides a complete and 
simple description of adiabatic spin 
pumping through the device in the limit, 
$\sqrt{t^2+\alpha^2} \gg \Gamma$, and $T\to0$.  In practice, however, this projective approach turns out to be 
a reliable approximation under much weaker conditions: we verified  in the noninteracting case $U=0$Ê
 that even for $\sqrt{t^2+\alpha^2} \gtrsim \Gamma$ it reproduces the exact results of Ref.~\cite{Brosco}  
 for the pumping fields with a few percent accuracy, and there is no reason why this accuracy should be decreased in the presence of strong 
 interactions in the mixed valence regime, the focus of our interest. The projective approach is thus able to approach the 
 regime $\Gamma \sim \alpha$ where the strongest spin pumping is expected~\cite{Brosco}.

\begin{figure}[t]
\includegraphics[width=0.75\columnwidth, clip]{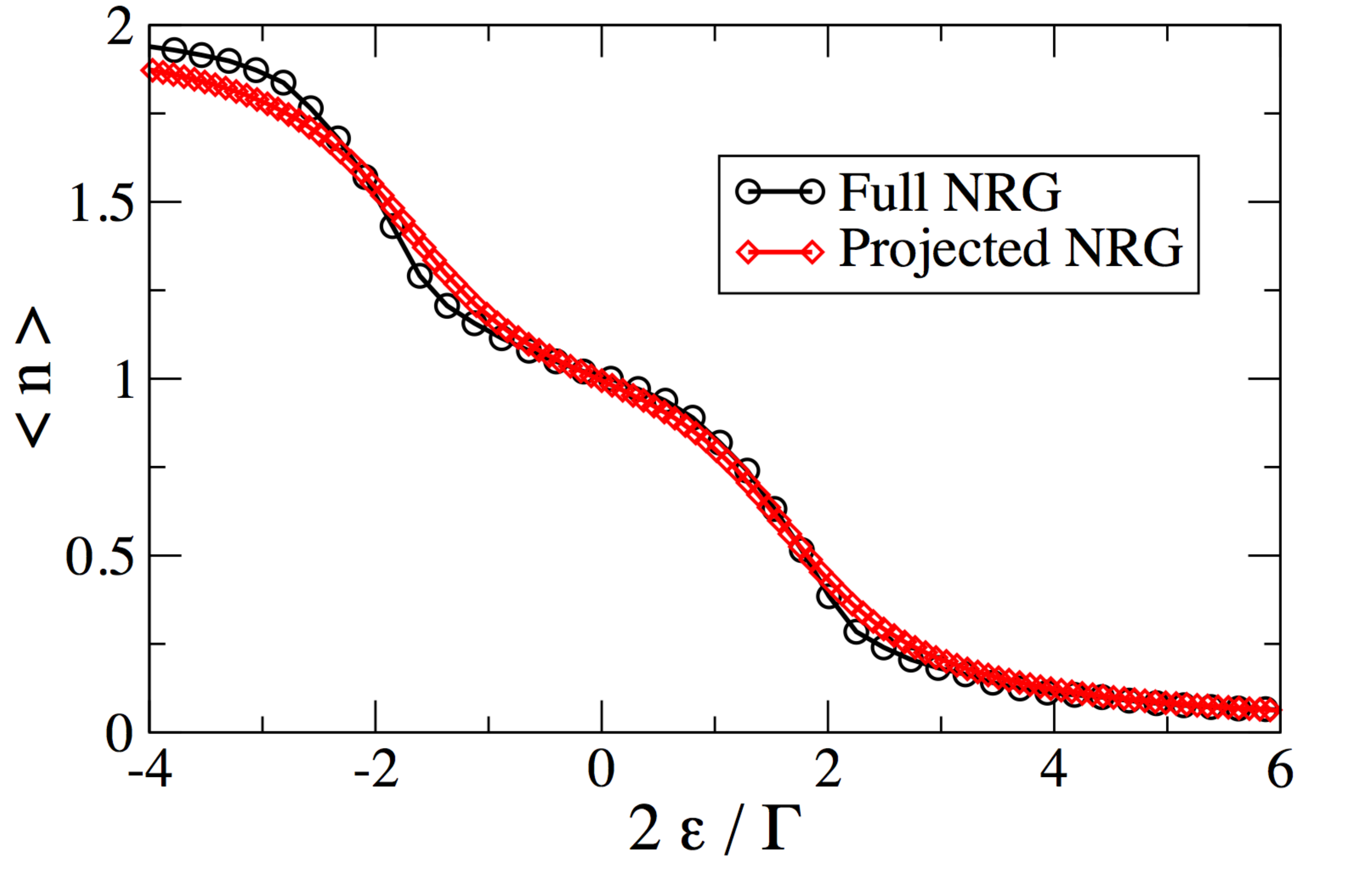}
\caption{\label{fig:nrg_frg_occupation} (Color online) 
Comparison of the projected NRG and full DM-NRG results for the total 
occupation of the dot for $\e =\e_1\equiv \e_2 $ for  couplings
 $v_{1}^{L} = 0.33\,\sqrt{\Gamma}$, 
 $v_{1}^{R} = 0.37\, \sqrt{\Gamma}$, 
 $v_{2}^{L} = 0.42\, \sqrt{\Gamma}$
and $v_{2}^{R} = -0.46\,\sqrt{\Gamma}$. 
The  Coulomb energy is $U = 2\, \Gamma$, $\alpha=0.9\, \Gamma$ and $t = 0.2\, \Gamma$.
\label{fig:occupation1}}
\end{figure}

\begin{figure}[t]
\includegraphics[width=0.8\columnwidth, clip]{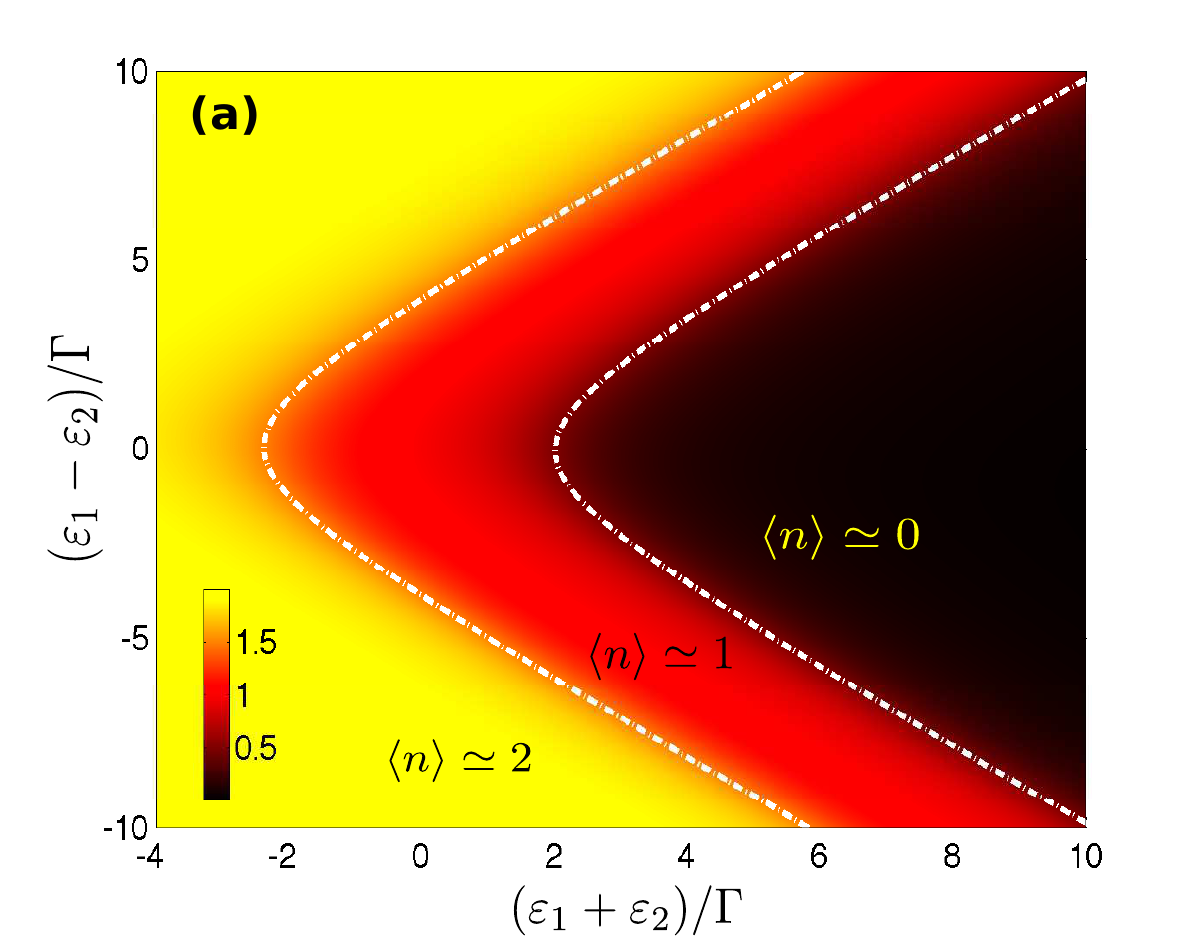}  
\includegraphics[width=0.8\columnwidth, clip]{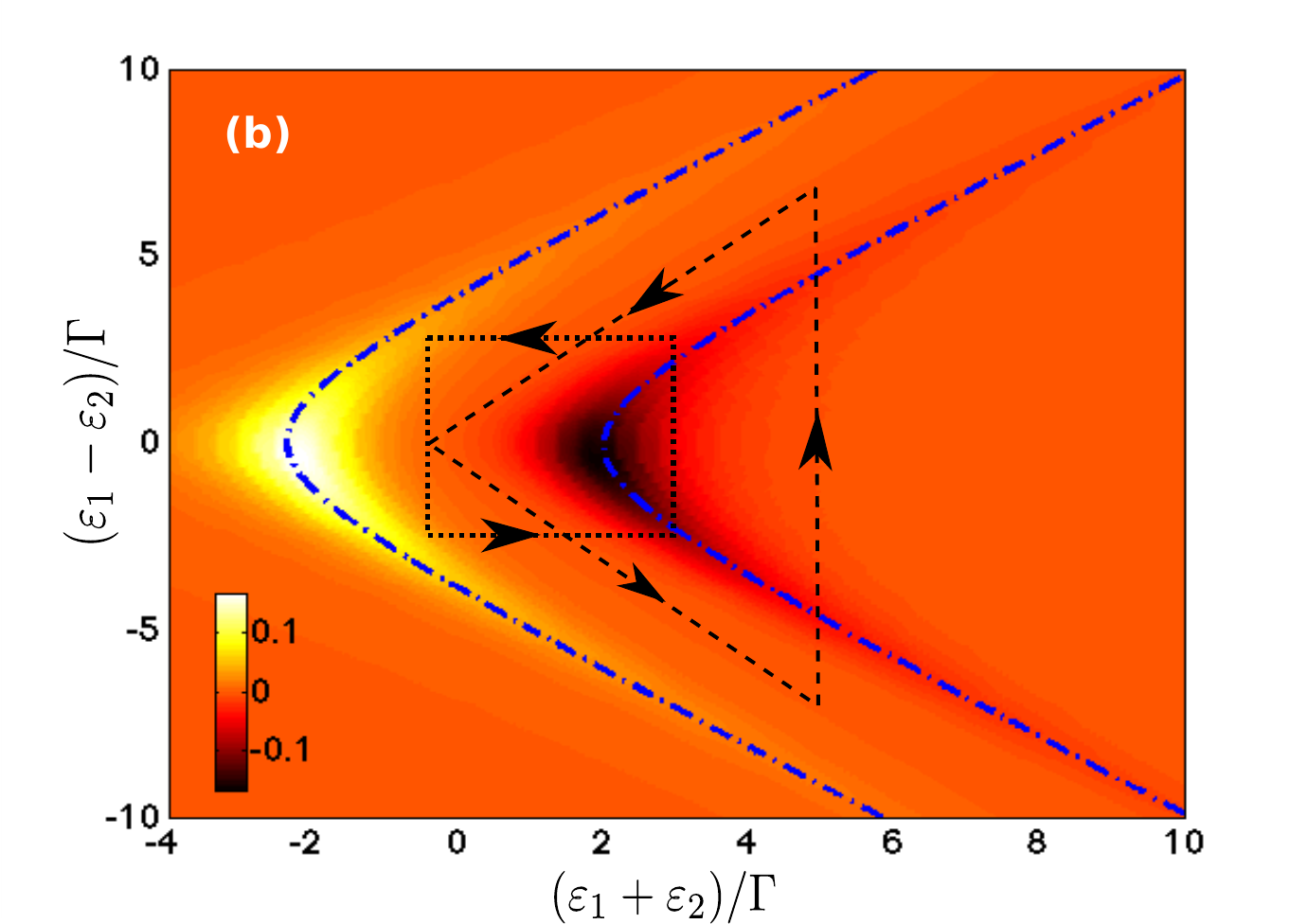}  
\caption{\label{fig:nrg_frg_occupation} (Color online) (a)
Density plot for the total occupation number in the 
$[\bar \e, \Delta \e ]$ plane for  the same parameters as in Fig.~\ref{fig:occupation1}. 
(b) Dimensionless  spin field 
$\Gamma^2\, \mathbf\Pi_L^{(S)}(\bar \e, \Delta \e)$ for the same parameters. 
The dashed black triangle indicates the pumping cycle used in Fig.~ \ref{fig:averages_orbital}.
Fig.~\ref{fig:sketch} shows the cycle indicated by dotted black lines. 
Dash-dotted blue lines denote the mixed valence regimes, 
where the total occupation is  $\langle n\rangle =0.5$  ($\langle n\rangle=1.5$).
\label{fig:occupation}}
\end{figure}

\emph{ Results:} To compute  the pumping fields, we employed  the density matrix NRG (DM-NRG)
approach~\cite{BudapestCode} to compute $\langle n\rangle$, and exploited 
the Friedel sum rule to construct   $G_{D} (\omega=0)$  and the S-matrix
as a function of  $\varepsilon_1$ and $\varepsilon_2$ using Eq.~\eqref{eq:Ssimple}. 
To check the validity of our projective approach, we also performed DM-NRG calculations for the unprojected 
Hamiltonian and determined the occupation $\langle n\rangle $~\footnote{These  calculations  are numerically expensive, since only U(1) and $Z_2$ symmetries can be used}. The agreement is very good 
(see Fig.~\ref{fig:occupation1}): the location as well as the shape  of the charging steps are
reproduced accurately  by  the projected Hamiltonian.

In Fig.~\ref{fig:occupation}, we present the spin pumping field, Eq.~\eqref{eq:field}, as well as 
the occupation $\langle n \rangle$ as a function of $\bar \varepsilon$ and  $\Delta\varepsilon$.
Two strong  resonances appear for $\Delta \varepsilon \approx 0 $, 
precisely in the vicinity of the  mixed valence regimes. The first resonance 
at  $\bar \epsilon\approx \sqrt{\alpha^2 + t^2}$ corresponds to the $n=0\leftrightarrow 1$ transition, and resembles  very much to the 
resonance found in the non-interacting case~\cite{Brosco}. 
Encircling this  first resonance  corresponds to a cycle sketched in Fig.~\ref{fig:sketch}: (1) first one populates level $\varepsilon_1$ by pulling it below the Fermi level. Then, (2) exchanging 
$\varepsilon_1\leftrightarrow \varepsilon_2$ one changes the spin content of the lower level, $E_-$. (3) Finally, 
one empties the level by pulling it over the Fermi energy. 

However, a surprising  second antiresonance appears at  $\bar \epsilon\approx \sqrt{\alpha^2 + t^2}-U$.  
This antiresonance is associated with the transition $n=1\leftrightarrow 2$. It
emerges solely as  a consequence of strong Coulomb interactions, and cannot be explained 
within a non-interacting picture. It "mirrors" the first resonance, but it carries just the opposite spin. 
This can be intuitively understood as follows: The doubly occupied level is a Kramers singlet and carries no spin. 
Therefore, the second electron entering the quantum dot must carry a spin 
opposite to the first one. 

\begin{figure}[t]
\begin{center}
\includegraphics[width= 0.8\columnwidth, clip]{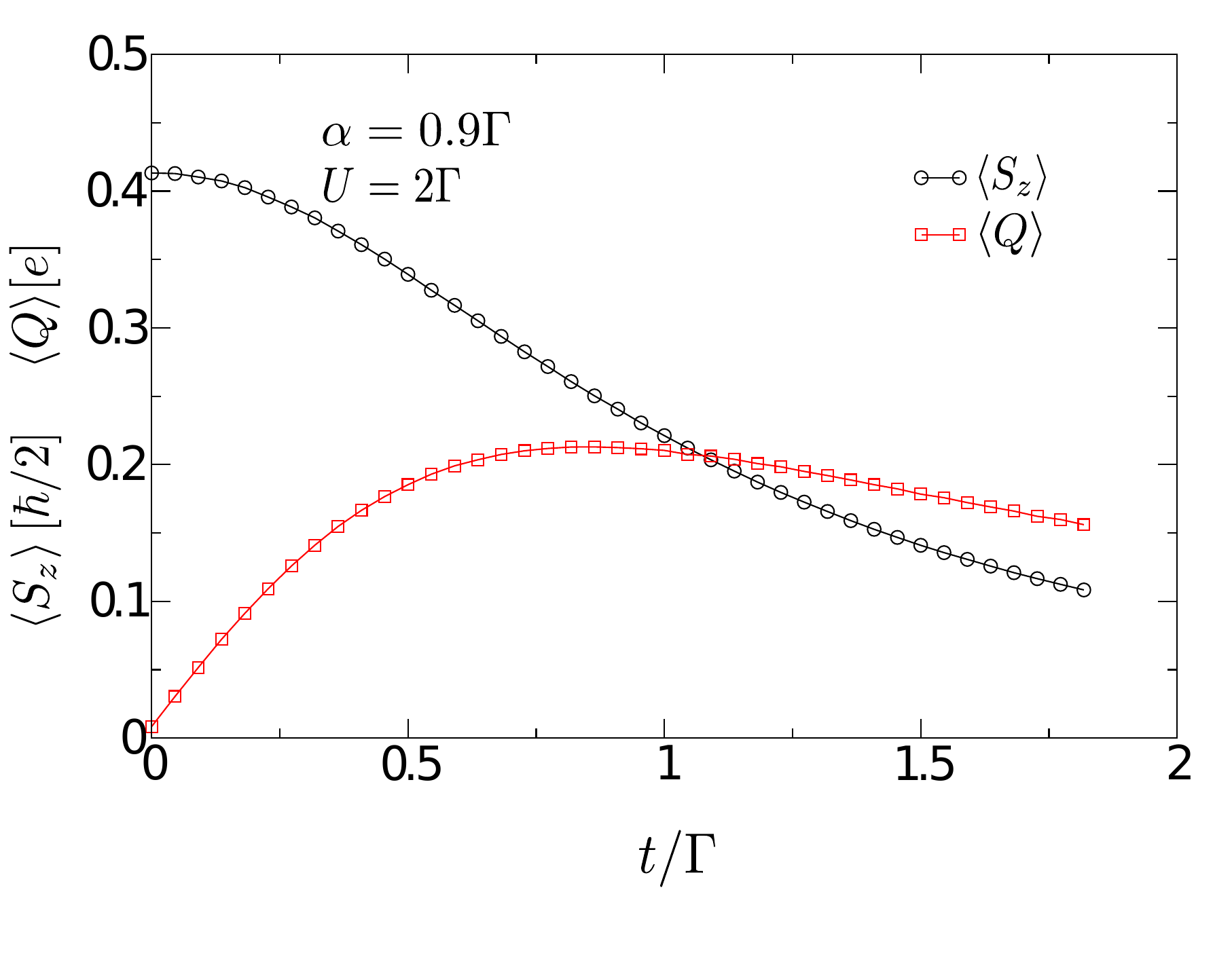}
\end{center}
\caption{\label{fig:averages_orbital}
Pumped charge and spin per cycle as function of the hybridization $t$ of the two levels, as computed 
for the triangle-shaped cycle in Fig.~\ref{fig:nrg_frg_occupation}.b.  The couplings are fixed to 
 $v_{1}^{L} = 0.33\,\sqrt{\Gamma}$, 
 $v_{1}^{R} = 0.37\, \sqrt{\Gamma}$, 
 $v_{2}^{L} = 0.42\, \sqrt{\Gamma}$
and $v_{2}^{R} = -0.46\,\sqrt{\Gamma}$. $\langle S_z\rangle$ is measured
in units of $\hbar/2$ and the pumped charge $\langle Q \rangle$ in units of e. 
 }
\end{figure}

To characterize the strength of the observed resonances, we computed the total spin pumped through a cycle, 
$(\varepsilon_1,\varepsilon_2)=(0,0)\to (0,5\, \Gamma)\to (5\,\Gamma,0)\to (0,0)$ (triangle in Fig.~\ref{fig:occupation}).  
For optimal parameters, 
the total spin pumped can reach values of the order of   $\sim \hbar/2$. The value of the pumped spin is almost independent of the Coulomb interaction as long as $U$ is sufficiently large. However, since the pumping originates from the large 
amplitude of the spin flip process during the avoided level crossing at $\varepsilon_1\approx\varepsilon_2$, its strength 
 is relatively sensitive to the spin independent interlevel hybridization, $t$, which suppresses the amplitude of these spin flip processes, 
and gradually suppresses the pumped spin (see Fig.~\ref {fig:averages_orbital}).

\begin{figure}[t]
\begin{center}
\includegraphics[width= 1\columnwidth, clip]{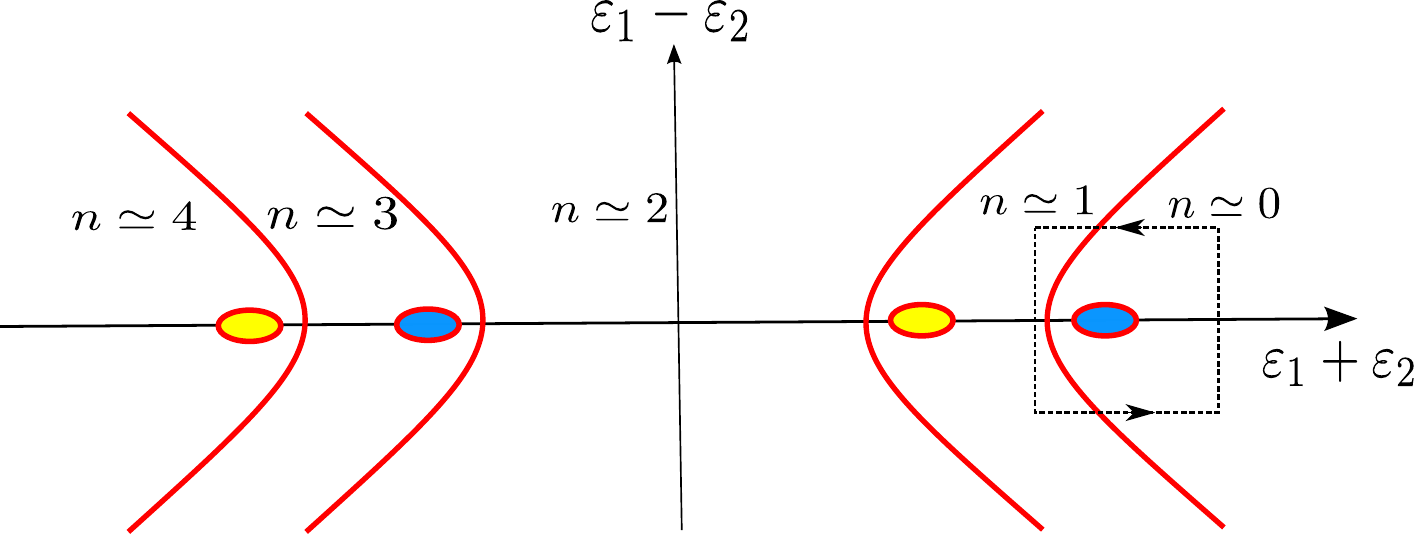}
\end{center}
\caption{\label{fig:phases_sketch}	
(Color online)  Sketch of the occupation of the dot and the position of the 
spin pumping resonances.}
\end{figure}

Our  projective approach can easily   be extended to the regime $\langle n \rangle \ge 2$ by means of an electron-hole transformation, which symmetry also allows us to determine the structure of the pumping fields in the whole parameter region (see Fig.~\ref{fig:phases_sketch}):  altogether we find \emph{two pairs of } spin pumping resonances, two resonances corresponding to the charging of  each Kramers degenerate level. 

For generic couplings, $v_j^r$, spin pumping is also accompanied by charge pumping, which, however,  may be strongly suppressed 
for special geometries. For a symmetrical device, e.g., with 
$v_1^L=v_1^R$ and $v_2^L=-v_2^R$ the charge field vanishes identically, and one obtains pure spin pumping, 
similar to the non-interacting case~\cite{SanJose}.  

\emph{Conclusions:} In the present paper, we used the concepts of Fermi liquid theory to formulate 
low temperature spin pumping through an interacting many-body system in terms  of the many-body S-matrix. 
We applied this formalism for a strongly interacting quantum dot with two gate-tuned levels, and showed that 
-- due to the strong interactions -- pumping field strong resonances and anti-resonances appear at every mixed valence 
transition, which can be  used to pump purely electronically a spin of the order $\sim \hbar/\mathrm{cycle}$ in a controlled way. 

\emph{ Acknowledgments.}
We acknowledge useful discussions with Sylvia Kusminskiy and J\"urgen K\"onig.  This research has been
supported by Hungarian Research Funds under grant Nos. 
 K105149, CNK80991, TAMOP-4.2.1/B-09/1/KMR-2010-0002, 
 by the 
UEFISCDI under French-Romanian Grant
DYMESYS (ANR 2011-IS04-001-01 and Contract
No. PN-II-ID-JRP-2011-1) and the EU-NKTH
GEOMDISS project.

\bibliography{references}

\end{document}